\documentclass[aps,prb,twocolumn,a4paper,10pt,showkeys,showpacs,preprintnumbers,floatfix,superscriptaddress]{revtex4-1}

\usepackage{graphicx}		
\usepackage{dcolumn}		
\usepackage{bm}				
\usepackage{amsmath}
\usepackage{amssymb}
\usepackage{amsfonts}
\usepackage{times}
\usepackage{tabularx}
\usepackage{xcolor,colortbl}
\usepackage{mathtools}

\definecolor{lightgray}{gray}{0.9}

\DeclarePairedDelimiterX\braket[2]{\langle}{\rangle}{#1 \delimsize\vert #2}

\begin{document}

\def\mean#1{\left< #1 \right>}

\title{A quantum interference capacitor based on \\ double-passage Landau-Zener-St\"{u}ckelberg-Majorana interferometry}

\author{Rub\'{e}n M. Otxoa}
\affiliation{Hitachi Cambridge Laboratory, J. J. Thomson Avenue, CB3 OHE, Cambridge, United Kingdom}
\affiliation{Donostia International Physics Center and Centro de F\'{i}sica de Materiales CSIC-UPV/EHU, Donostia-San Sebasti\'{a}n 20018, Spain}
\author{Anasua Chatterjee}
\affiliation{London Center for Nanotechnology, University College London, London, WC1H 0AH, United Kingdom}
\author{Sergey N. Shevchenko}
\affiliation{B. Verkin Institute for Low Temperature Physics and Engineering, Kharkov 61103, Ukraine}
\affiliation{V. N. Karazin Kharkov National University, Kharkov 61022, Ukraine}
\affiliation{Theoretical Quantum Physics Laboratory, Cluster for Pioneering
	Research, RIKEN, Wako-shi, Saitama 351-0198, Japan}
\author{Sylvain Barraud}
\affiliation{CEA/LETI-MINATEC, CEA-Grenoble, 38000 Grenoble, France}
\author{Franco Nori}
\affiliation{Theoretical Quantum Physics Laboratory, Cluster for Pioneering
	Research, RIKEN, Wako-shi, Saitama 351-0198, Japan}
\affiliation{Physics Department, University of Michigan, Ann Arbor, MI 48109-1040, USA}
\author{M. Fernando Gonzalez-Zalba}
\affiliation{Hitachi Cambridge Laboratory, J. J. Thomson Avenue, CB3 OHE, Cambridge, United Kingdom}

\date{\today}

\begin{abstract}

The implementation of quantum technologies in electronics leads naturally to the concept of coherent single-electron circuits, in which a single charge is used coherently to provide enhanced performance.  In this work, we propose a coherent single-electron device that operates as an electrically-tunable  capacitor. This system exhibits a sinusoidal dependence of the capacitance with voltage, in which the amplitude of the capacitance changes and the voltage period can be tuned by electric means. The device concept is based on double-passage Landau-Zener-St\"{u}ckelberg-Majorana interferometry of a coupled two-level system that is further tunnel-coupled to an electron reservoir. We test this model experimentally by performing Landau-Zener-St\"{u}ckelberg-Majorana interferometry in a single-electron double quantum dot coupled to an electron reservoir and show that the voltage period of the capacitance oscillations is directly proportional to the excitation frequency and that the amplitude of the oscillations depends on the dynamical parameters of the system: intrinsic relaxation and coherence time, as well as the tunneling rate to the reservoir. Our work opens up an opportunity to use the non-linear capacitance of double quantum dots to obtain enhanced device functionalities. 



\end{abstract}

\maketitle

\section{Introduction}

The new wave of quantum technologies aims at using basic principles of quantum mechanics, such as superposition or entanglement, to obtain functionality beyond what conventional devices can provide~\cite{Zagoskin2011, Gu2017, Acin2018}. In the field of nanoelectronics, superposition and entanglement can be harnessed to build coherent quantum circuits that can be used, for example, for quantum information processing~\cite{Acin2018}, precision sensing~\cite{Degen2017} and quantum-limited amplification~\cite{Roy2016, SchaalJPA2019}. To produce a coherent superposition between quantum states in nanoelectronic circuits, Landau-Zener-St\"{u}ckelberg-Majorana (LZSM) interferometry~\cite{nakamura2012nonadiabatic, Shevchenko2019} is a prime example. In LZSM interferometry, a quantum two-level system~\cite{Buluta2011} is driven strongly across an avoided energy-level crossing producing first a quantum superposition between the ground and excited state of the system. These states evolve with different dynamical phases and, following a second passage through the anticrossing, coherent interference between these two states can occur~\cite{shytov2003landau, shevchenko2010}. LZSM interference has been observed in a number of different platforms such as Rydberg atoms~\cite{rubbmark1981dynamical}, superconductive Josephson junctions~\cite{shevchenko2010,oliver2005mach,berns2006},  nitrogen vacancy centres in diamond~\cite{fuchs2011quantum}, silicon charge qubits in complementary metal-oxide semiconductor (CMOS) technology~\cite{Stehlik2012, Dupont-Ferrier2013, gonzalez2016gate, Ono2019, Ono2019b} and silicon carbide devices~\cite{Miao2019}. Moreover, it has been used as a diagnostic tool to obtain physical parameters of two-level systems as well as a method for the fast manipulation of spin-based qubits~\cite{Petta2010}.

Although standard LZSM interferometry has been extensively studied for two-level systems, realistic quantum systems may have more than just two levels. Multi-level LZSM physics has been observed, for example, in superconducting qubits~\cite{Berns2008, Wang2010, Chen2011, Ferron2016, Silveri2017} and semiconductor quantum dots~\cite{Reynoso2012, Shi2014, Stehlik2016, Chen2017, Qi2017, Chatterjee2018, Shevchenko2018, Pasek2018}.

Here, we present an application of multi-level LZSM interferometry to demonstrate a novel device that presents a sinusoidal dependence of the capacitance with voltage in which the amplitude of the capacitance changes, and the voltage period can be tuned by electric means. The device is a tunable capacitor based on LZSM interferometry of a coupled two-level system that is further tunnel coupled to an electron reservoir. In the double-passage regime, we find that the capacitance of the system varies periodically with the bias voltage and that the amplitude of the capacitance changes depends on the intrinsic relaxation and phase coherence times of the electron as well as the tunnel rate to the reservoir. We implement the capacitor experimentally using a silicon single-electron double quantum dot defined in the top-most corners of a nanowire transistor~\cite{Voisin2014, Ibberson2018} that is also coupled to an electronic reservoir. We drive the system in the LZSM double-passage regime using microwave excitations and probe the non-linear parametric capacitance of the driven system using radiowave reflectometry. Finally, we compare the theory and experiment and find good agreement that enables us to determine the dynamical parameters of the system: intrinsic relaxation and coherence time as well as quantum-dot-reservoir tunneling rate.

\begin{figure*}[t]
	\includegraphics{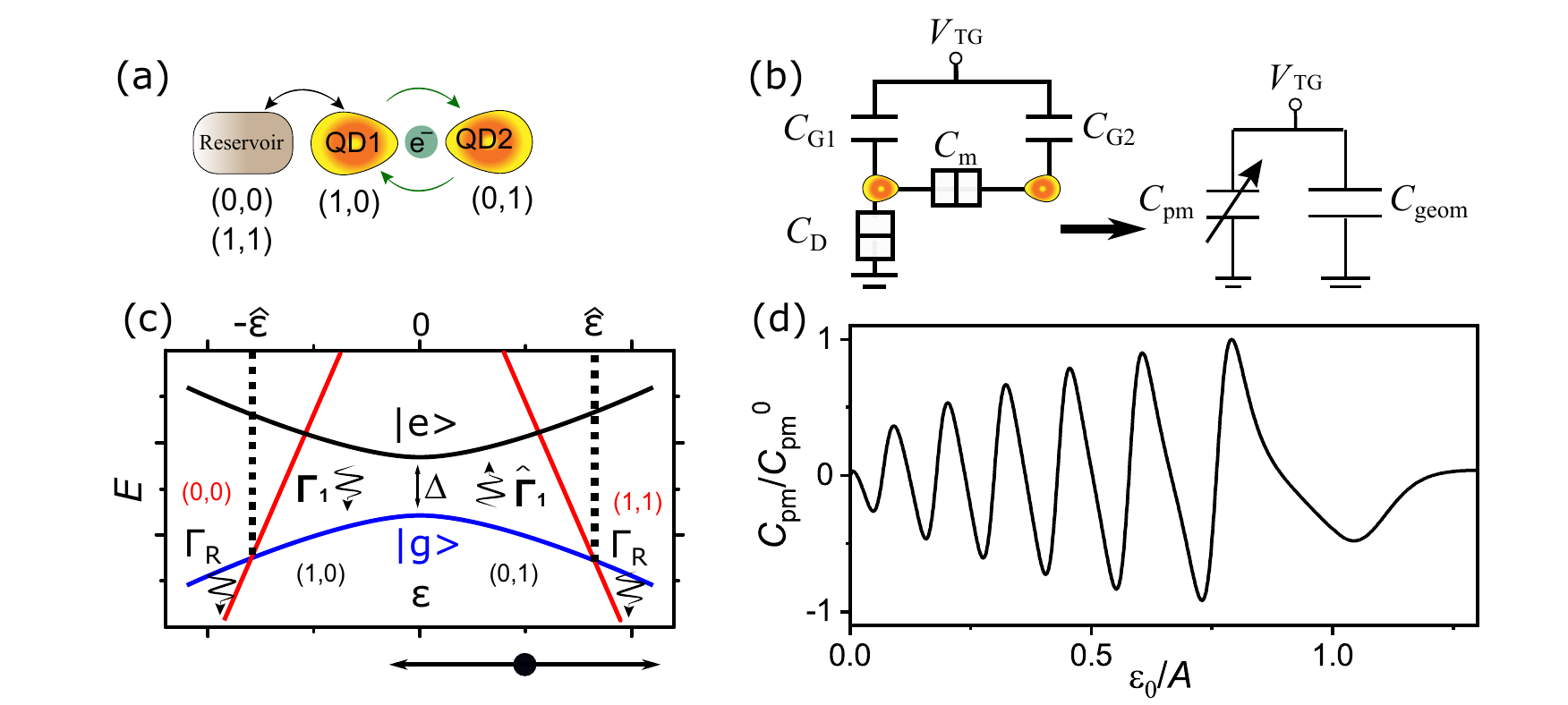}
	\caption{\label{Figure1} Quantum interference capacitor proposal. (a) Schematic illustration of a DQD coupled to a reservoir and the relevant charge states and tunneling processes. (b) Left, circuit representation of the DQD coupled to a reservoir. Right, equivalent circuit which includes the geometrical $C_{\text{geom}}$ and parametric $C_{\text{pm}}$ capacitance in parallel. (c) Energy levels of the DQD as a function of detuning. Here $\Gamma_{1}$ is the relaxation rate from $\left| e \right\rangle $ to $\left| g \right\rangle $, and $\hat{\Gamma}_{1}$ from $\left| g \right\rangle $ to $\left| e \right\rangle $. The red lines indicate charge states involving the reservoir and cross $\left| g \right\rangle $ at $\varepsilon=\pm\hat{\varepsilon}$. $\Gamma_\text{R}$ indicates the QD1-reservoir tunnel rate. Below the graph, the horizontal double-arrow indicates the cycle to perform double-passage LZSM interferometry and its central dot indicates the offset detuning. (d) Simulated normalized parametric capacitance $C_\text{pm}/C_\text{pm}^0$ versus reduced detuning $\varepsilon_0/A$, for $T_{\text{1}}=50$~ns, $T_{2}=35$~ps, $T_\text{R}=30$~ps, and $\omega/2\pi=11$\,GHz.}
\end{figure*}

\section{Quantum Interference Capacitor proposal}

In this section, we describe the physical requirements of the quantum interference capacitor. We consider a quantum two-level system where the two levels correspond to two different charge states. The energy difference between levels can be controlled by a tuning parameter $\varepsilon$ in timescales comparable or faster than the characteristic relaxation ($T_1$) and coherence times ($T_2$) of the system. The two levels are coupled via a coupling term $\Delta$. The Hamiltonian of the coupled two-level system, expressed in terms of the Pauli spin matrices, is

\begin{equation}
\label{hamiltonian}
\begin{split}
H(t)=-\frac{\Delta}{2}\sigma_{x}-\frac{\varepsilon(t)}{2}\sigma_{z}.
\end{split}
\end{equation}

Furthermore, the system must be tunnel-coupled to the charge reservoir to allow particle exchange. These elementary requirements can be found in a variety of systems~\cite{Buluta2011}, such as superconducting charge qubits~\cite{}, impurities in semiconductors~\cite{Dupont-Ferrier2013} and double quantum dots (DQDs)~\cite{Bogan2018, Koski2018}. In this paper, we focus in the latter for the case where the charged particles are electrons.

In a single-electron DQD, an electron is shared among the QDs giving rise to two possible classical charge configurations $(n_{1}n_{2})=(10)$ and $(01)$, where $n_i$ corresponds to the number of charges in the $i$th QD. We consider the case in which an electron can tunnel between the QDs and can also exchange particles with an electron reservoir (states (00) or (11)), see Fig.~\ref{Figure1}(a). For a DQD, $\varepsilon$ represents the energy detuning between the $(10)$ and $(01)$ charge states, and $\Delta$ is the tunnel coupling that mixes them at $\varepsilon=0$.

In Fig.$\,$\ref{Figure1}(b), we present the minimal electrical circuit to implement the quantum interference capacitor. The two QDs, are connected to a top-gate electrode via the gate capacitances $C_{\text{G}i}$. The QDs are tunnel coupled to each other via a mutual capacitance $C_\text{m}$, and QD1 is further tunnel-coupled to a reservoir via a capacitance $C_\text{D}$. The differential capacitance, as seen from the top-gate, can be expressed~\cite{mizuta2017,Esterli2019} as 

\begin{equation}
	C_\text{diff}=e\frac{\partial (n_1+n_2)}{\partial V_\text{TG}}=C_\text{geom}+C_\text{pm},
\end{equation}

\noindent where $e$ is the electron charge and $V_\text{TG}$ is the top-gate voltage. Here $C_\text{geom}$ is the geometrical capacitance  and $C_\text{pm}$ is a voltage-dependent term, the parametric capacitance, see the equivalent circuit on the right side of Fig.~\ref{Figure1}(b). We consider the weak coupling limit $C_\text{m}\ll C_{\text{G}i}, C_\text{D}$, where the geometrical capacitance reads $C_\text{geom}=C_{\text{G}2}/(C_{\text{G}2}+C_\text{D})$. The parametric capacitance can be probed with a sinusoidal detuning $\varepsilon(t)=\varepsilon_0+\varepsilon_\text{r}\text{sin}(\omega_\text{r} t)$ and, when it has a small amplitude $\varepsilon_\text{r}\ll\Delta$, low frequency ($\omega_\text{r}$ lower than the relaxation rates of the system) and offset $\varepsilon_0$, its average value can be expressed as~\cite{Chatterjee2018},

\begin{equation}
\label{Cp}
C_\text{pm}=2e^{2}\alpha_{-}^2\frac{\partial}{\partial\varepsilon_0}\left[P_{01}-P_{10}+\frac{\alpha_{+}}{\alpha_{-}}(P_{00}+P_{11})\right].
\end{equation}

Here $\alpha_{\pm}=\left(\alpha_{2}\pm\alpha_{1}\right)/2$, where $\alpha_1=C_{\text{G}1}/(C_{\text{G}1}+C_\text{m})$ and $\alpha_2=C_{\text{G}2}/(C_{\text{G}2}+C_\text{m}+C_\text{D})$ are the QD-gate couplings. Note that we have used the expression $\varepsilon=-2e\alpha_{-}\left(V_{\text{TG}}-V_{\text{TG}}^{0}\right)$, with $V_{\text{TG}}^{0}$ denoting the top-gate voltage where the states $(10)$ and $(01)$ anticross, to relate the top-gate voltage to the induced detuning. Finally, $P_{n_{1}n_{2}}$ refers to the probability of being in the electronic state $(n_{1}n_{2})$. For the implementation of the quantum interference capacitor, we will consider DQDs with similar gate couplings, $\alpha_-\ll\alpha_+$, so that the parametric capacitance in Eq.~(\ref{Cp}) is predominately determined by changes in $P_{00}$ or $P_{11}$.

Next, we subject the DQD to a faster oscillatory detuning $\varepsilon(t)=\varepsilon_{0}+A\text{sin}(\omega t)+\delta\varepsilon(t)$, where $A$ is the amplitude of the detuning oscillations, $\omega$ is the frequency of the driving field ($\omega\gg\omega_\text{r}$) and $\delta\varepsilon(t)$ is the classical noise~\cite{}. When a coupled two-level system is subject to periodic driving with sufficiently large amplitude, LZSM transitions between the ground $\left| g \right\rangle $ and excited state $\left| e \right\rangle $ of the Hamiltonian in Eq.~(\ref{hamiltonian}) can occur. We consider the scenario in which the system performs a double-passage through the anticrossing producing the LZSM interference and then a QD exchanges particles with the electron reservoir, see Fig.~\ref{Figure1}(c). For simplicity, we explain the cycle that involves the (01)-(11) particle exchange process (indicated by the black horizontal arrow) although the discussion also applies for the symmetric drive with respect to $\varepsilon=0$ where the exchange is (00)-(10).


The dynamics of the two-level system can be described by a master equation:

\begin{equation}
\label{Master}
\begin{split}
\partial_{t}P_{g}=\left[W(\varepsilon_0)+\Gamma_1(\varepsilon_0)\right]P_{e}-\left[W(\varepsilon_0)+\hat{\Gamma}_1(\varepsilon_0)\right]P_{g},\\
P_{g}+P_{e}=1
\end{split}
\end{equation}


\noindent where $W$ is the rate of the LZSM transitions, $\Gamma_{1}$ is the relaxation rate from the excited state $\left| e \right\rangle $ to the ground state $\left| g \right\rangle $, and $\hat{\Gamma}_{1}$ from $\left| g \right\rangle $ to $\left| e\right\rangle $, see Fig.~\ref{Figure1}(c). We consider the low-temperature limit $k_\text{B}T\ll\Delta$, where $\Gamma_{1}=1/T_1$ and $\hat{\Gamma}_{1}=0$. Therefore, Eq.$\,$(\ref{Master}) can be written as

\begin{equation}
 \partial_{t}P_{g}=\left[W(\varepsilon_0)+\Gamma_{1}\right]P_{e}-W(\varepsilon_0)P_{g}.
 \end{equation}
 
 We calculate the stationary solution of the system and find

\begin{equation}
\label{prob_e}
\begin{split}
P_{g}=1-\frac{W(\varepsilon_0)}{2\,W(\varepsilon_0)+\Gamma_{1}}.
\end{split}
\end{equation}

After a second passage, considering (01) as a starting point, the system exchanges electrons with the reservoir. The probability $P_{11}$ at that point can be expressed as

\begin{equation}
\label{prob_e2}
\begin{split}
P_{11}=P_{\text{R}}(\varepsilon_0)\left(1-\frac{W(\varepsilon_0)}{2\,W(\varepsilon_0)+\Gamma_{1}}\right),
\end{split}
\end{equation}

\noindent where $P_{\text{R}}$ represents the tunneling probability to the reservoir. $P_\text{R}$ can be expressed as $P_{\text{R}}(\varepsilon_0)=1-\exp{(-t_\text{R}/T_\text{R})}$, where $T_{\text{R}}$ is the QD-reservoir relaxation time and $t_\text{R}$ represents the time the electron spends after passing the crossing point between the (01) and (11) charge states at $\varepsilon=\hat{\varepsilon}$. Given the functional shape of the drive, we obtain

\begin{equation}
\label{RateEquation}
\begin{split}
P_{\text{R}}(\varepsilon_0)=1-\text{exp}\left\{\frac{-1}{T_\text{R}\omega}\left[\pi-2\,\arcsin\left(\frac{\hat{\varepsilon}-\varepsilon_0}{A}\right)\right]\right\}.
\end{split}
\end{equation}

The probability of the (11) state increases as the system expends more time passing the crossing point. Eventually, we calculate the derivative of the probability $P_{11}$ with respect to the detuning that enters in Eq.~(\ref{Cp}),

\begin{equation}
\label{MasterEq.Final}
\begin{split}
\partial_{\varepsilon_0}P_{11}=\partial_{\varepsilon_0}P_{\text{R}}(\varepsilon_0)\left[1-\frac{W(\varepsilon_0)}{2\,W(\varepsilon_0)+\Gamma_1}\right]-\\
-P_{\text{R}}(\varepsilon_0)\frac{\Gamma_1\partial_{\varepsilon_0}W(\varepsilon_0)}{\left[2\,W(\varepsilon_0)+\Gamma_1\right]^2}.
\end{split}
\end{equation}

As the detuning gets closer to $\hat{\varepsilon}$, the first term in Eq.$\,$(\ref{MasterEq.Final}) becomes negligible compared to the second one, leading to the final expression for the variation of the probability $P_{11}$:

\begin{equation}
\label{MasterEq.Final_1}
\begin{split}
\partial_{\varepsilon_0}P_{11}\approx-P_{\text{R}}(\varepsilon_0)\frac{T_1\partial_{\varepsilon_0}W(\varepsilon_0)}{\left[1+2\,W(\varepsilon_0) \text{T}_{1}\right]^2}.
\end{split}
\end{equation}


\begin{figure*}[t]
	\includegraphics{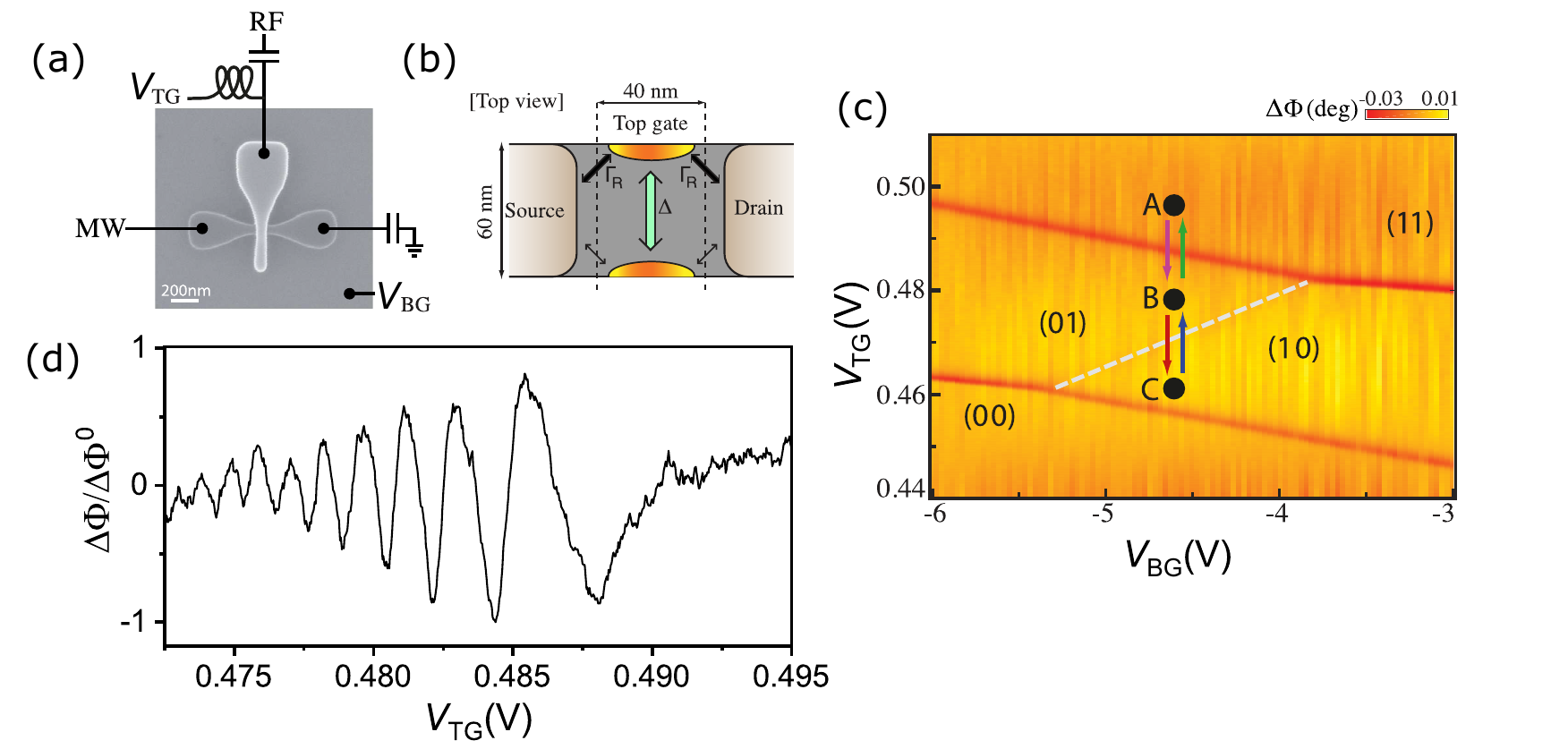}
	\caption{\label{Figure2} Experiment. (a) Scanning electron microscope image of a device similar to the one measured connected to a radio frequency reflectometry set-up via the top gate. $V_\text{TG}$ is applied via an on-chip bias tee with a 100~pF capacitor and 560~nH inductor. The drain of the device is AC grounded via a 100~pF capacitor. (b) Schematic of the device indicating the location of the corner quantum dots in top-view, with the top gate transparent for clarity. The electronic transitions are marked by arrows, and $\Delta$ represents the tunnel coupling. (c) Colour map of the DQD charge stability diagram extracted from reflectometry measurements. The dashed white line indicates the interdot charge transition. The letters indicate s
uential operations: starting from the (11) state at point A, followed by the unloading of an electron to (10) at point B, the creation of a superposition at point C, a return back to B across the transition to create the interference, which is then projected back onto (11) when returning to A. (d) Measured normalised resonator phase response vs gate voltage, $V_{\text{TG}}$, for a probing frequency $\omega/2\pi=11$ GHz.}
\end{figure*}

Therefore, the problem of calculating $C_\text{pm}$ reduces to calculating the rate of the LZSM transitions, which we do in the following. After the first passage through the anticrossing, assuming $\hbar=1$, the system acquires a dynamical phase due to the energy difference between the two energy states as follows \cite{berns2006}

\begin{equation}
\label{phase}
\begin{split}
\Delta\theta(\tau)=\int\limits_{t}^{t+\tau}(E_{e}-E_{g})dt= \Delta \int\limits_{t}^{t+\tau} e^{-\mathrm{i}\phi(\text{t})}dt,
\end{split}
\end{equation}

\noindent where $\phi(t)$ refers to the driving mechanism, $\phi(\tau)=\int\limits_{\text{0}}^{\tau}\varepsilon(t)dt$. Once the system is far from the avoided crossing, the $\left| e \right\rangle $ and $\left| g \right\rangle $ states evolve independently, accumulating the so-called St\"{u}ckelberg phase, $\Delta\theta_{e \leftrightarrow g}$ and the rate of LZSM transitions can be expressed as

\begin{equation}
\label{}
\begin{split}
 W(\varepsilon)=\lim_{\tau\to\infty}\frac{\Delta\theta(t)_{e \leftrightarrow g}\Delta\theta^{*}(t+\tau)_{e \leftrightarrow g}}{\tau}.
\end{split}
\end{equation}

Using the Jacobi-Anger expansion, $\text{exp}(\mathrm{i}z\, \text{sin} \gamma)=\sum_{n={-\infty}^\infty} J_{n}(z)\text{e}^{\mathrm{i}n\gamma}$, where $J_{n}(z)$ are Bessel functions of the first kind. We associate the noise energy term $\exp{(-\mathrm{i}\delta\epsilon(t))}$ to a noise in the  phase $\exp{(-\mathrm{i}\delta\phi(t))}$. As a result, when we integrate Eq.\,(\ref{phase}), we obtain

\begin{equation}
\label{LZrate}
\begin{split}
W\left(\varepsilon_0\right)=\frac{\Delta^{2}}{2}\int\limits_{\text{0}}^{\tau}\sum_{n}J_{n}^{2}\left(\frac{A}{\omega}\right)\text{e}^{-t\left[\mathrm{i}T_2\left(\varepsilon_0-n\omega\right)-1\right]/T_2}dt,
\end{split}
\end{equation}\\

\noindent where we make use of the white noise theorem $\mean{\text{e}^{-\mathrm{i}\delta\phi(t)}\text{e}^{-\mathrm{i}\delta\phi(t+\tau)}}=\text{e}^{-\tau/T_{2}}$. Assuming $n$ to be large in Eq.\,(\ref{LZrate}), the Bessel function can be approximated to the Airy function as $J_{n}\left(\frac{A}{\omega}\right)=\frac{A}{\omega}\,\text{Ai}\left[\frac{A}{\omega}\left(n-\frac{A}{\omega}\right)\right]$. Using the approximation $\pi\,\text{cot}\,\pi z\approx\sum_{n=-\infty}^{\infty}\frac{1}{z-n}$, the LZSM transition rate becomes

\begin{equation}
\label{LZAiry}
\begin{split}
W\left(\varepsilon_0\right)=\frac{\pi\Delta^{2}\zeta^{2}}{2\omega}\text{Ai}^{2}\left[\frac{\zeta}{\omega}\left(\varepsilon_0-A\right)\right]\text{exp}^{-t_{1}/T_2},
\end{split}
\end{equation}

\noindent where 

\begin{equation}
\label{}
\begin{split}
\zeta=(2\, \omega/A)^{1/3}  \text{, and } t_{1}=2\left[\pi-\arcsin(\varepsilon_0/A)\right]/\omega 
\end{split}
\end{equation}

\noindent is the time after the first passage. We restore $\hbar$ and write the parametric capacitance explicitly:

\begin{equation}
\label{Cp_final}
\begin{split}
C_{\text{pm}}=\frac{2e^2\alpha_-\alpha_+\zeta}{\hbar\omega}\left[1-e^{-\frac{t_\text{R}}{T_\text{R}}}\right]\frac{\gamma\text{Ai}'\left[\frac{\zeta\left(\varepsilon_0-A\right)}{\hbar\omega}\right]\text{Ai}\left[\frac{\zeta\left(\varepsilon_0-A\right)}{\hbar\omega}\right]}{\left(1+\gamma\text{Ai}^{2}\left[\frac{\zeta\left(\varepsilon_0-A\right)}{\hbar\omega}\right]\right)^2},
\end{split}
\end{equation}



\noindent where 

\begin{equation}
\label{}
\begin{split}
\gamma=T_1\frac{\pi\zeta^{2}\Delta^{2}}{\hbar^2\omega}\exp(-t_1/T_2). 
\end{split}
\end{equation}

\noindent Using Eq.~(\ref{Cp_final}), in Fig.~\ref{Figure1}(d), we show a plot of the normalized $C_{\text{pm}}$ as a function of the reduced detuning $\varepsilon_0/A$ for $\hat{\varepsilon}=A$ and $\omega/2\pi=11$~GHz, $T_1=50$~ns, $T_2=35$~ps, and $T_\text{R}=30$~ps. For values $\varepsilon_0<A$, the parametric capacitance shows an oscillatory behaviour as a function of detuning, whereas for $\varepsilon_0>A$ the signal decays exponentially. In the oscillatory region, the variation of the amplitude of the oscillations with detuning is determined by $T_2$ and $T_\text{R}$, whereas the overall amplitude depends on $\omega$, $T_1$, $T_2$, and $T_\text{R}$. To facilitate the understanding of the functional dependence $C_{\text{pm}}$, in the limit $\varepsilon_0<A$, we find that Eq.~(\ref{Cp_final}) can be simplified to

\begin{equation}
\label{MasterEq.Final_2}
\begin{split}
C_{\text{pm}}\approx C_{\text{pm}}^{0}\left(\varepsilon_0, A,\omega,T_{\text{1}},T_{2},T_\text{{R}}\right)\text{cos}\left(2\pi V_\text{TG}/\delta \! V_\text{TG}\right),
\end{split}
\end{equation}

\noindent where $\delta \! V_\text{TG}= \pi\hbar\omega/(2\sqrt{2}e\alpha_-)$ is the top-gate voltage period. To elucidate the validity of our model, we study the implementation of the quantum interference capacitor using LZSM interferometry in a single-electron DQD strongly driven by a microwave (MW) field.

\begin{figure*}[t]
	\includegraphics{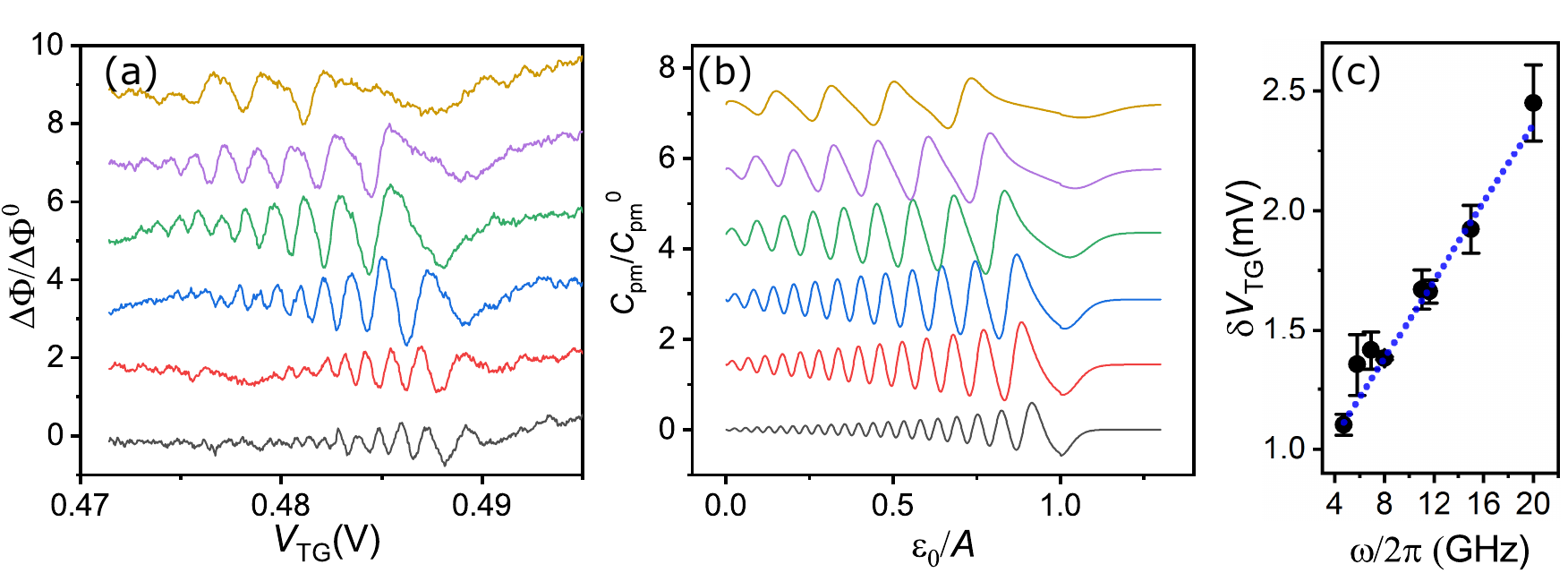}
	\caption{\label{Figure3} Microwave frequency dependence: experiment and theory. (a) Experimental normalized response of the resonator phase shift as a function of the top-gate voltage for microwave frequencies $\omega/2\pi=4.72$ (black), 6.9 (red), 8 (blue), 11 (green), 15 (purple), 21~GHz (yellow). Traces are displaced upwards by 1.8 for clarity.  (b) Calculated normalized parametric capacitance as a function of reduced detuning for the same frequencies as in (a) using $T_1=50$~ns, $T_2=35$~ps, and $T_\text{R}=30$~ps. Traces are displaced upwards by 1.45. (c) Position in $V_\text{TG}$ of the maximum of the Fourier transform of the experimental data in (a) as a function of $\omega$, and linear fit as theoretically expected from the Fourier transform of Eq.$\,$(\ref{MasterEq.Final_2}).
	}
\end{figure*}

\section{Experimental Results}

We now turn to a concrete implementation of the quantum interference capacitor, consisting of a single-electron DQD in which we perform LZSM interferometry. Our device consists of a silicon nanowire transistor fabricated using industrial 300~mm silicon-on-insulator (SOI) technology, as described in previous work~\cite{Voisin2014,Gonzalez-Zalba2015,Ibberson2018} and shown in Fig.$\,$\ref{Figure2}(a). The nanowire is 11~nm high, with a width of 60~nm, while a 40~nm wide wrap-around top-gate covers the nanowire, separated by a SiO$_2$/HfSiON dielectric layer. In such square-section transistors, when a positive top-gate voltage ($V_\text{TG}$) is applied, electron accumulation occurs along the top-most corners of the channel, resulting in a DQD in parallel with the source and drain electron reservoirs. This situation is shown in the schematic in Fig.$\,$\ref{Figure2}(b). The use of SOI technology enables back-gating the device by applying a voltage ($V_\text{BG}$) to the silicon intrinsic handle wafer, made temporarily conductive by flashing a blue LED placed on the sample printed circuit board at 35~mK.

To measure the parametric capacitance of the DQD, we employ radiofrequency (RF) reflectometry by interfacing the transistor with an electrical $LC$ resonator with a resonance frequency $\omega_\text{r}/2\pi=313\,$MHz and a loaded quality factor $Q\sim40$. The resonator is coupled to the DQD via the top gate for high-sensitivity dispersive readout~\cite{Colless2013,Ahmed2018,Pakkiam2018,West2019,Zheng2019} and consists of a surface mount inductor, $L=390$~nH, and the parasitic capacitance to ground of the device, $C_\text{p}=$~660~fF. Changes in device capacitance manifest as changes in resonant frequency ($\omega_\text{r}=1/\sqrt{L(C_\text{p}+C_\text{pm})}$) that we detect using low-noise cryogenic and room-temperature amplification, combined with homodyne detection. In order to perform LZSM interferometry, we apply MW signals directly on the source of the transistor. We operate the DQD in the charge qubit regime. The DQD is further coupled to an electron reservoir at the source and drain, with one quantum dot being significantly more coupled to the reservoirs than the other, as shown in Fig.$\,$\ref{Figure2}(b).

We measure the charge stability diagram of the device as a function of the top and back gates, as shown in Fig.$\,$\ref{Figure2}(c). The tunneling of single electrons produce changes in parametric capacitance, $C_\text{pm}$, that lead to changes in $\omega_\text{r}$. Since we measure at a single frequency, those changes appear as changes in the phase response of the reflected signal, $\Delta\Phi=-2QC_\text{pm}/C_\text{p}$. We see four stable charge configurations ($n_1n_2$). In the absence of additional charge transitions at lower gate voltages, we tentatively conclude that lower voltages result in the system being depleted of electrons. We therefore operate in the single-electron regime, with the electron occupying the left or right dot, denoted as the state (10) or (01). Loading or unloading of an electron from or into a reservoir leads to the states (11) and (00), respectively.

Next, we apply MWs (amplitude $A=\hat{\varepsilon}$ and frequency $\omega/2\pi=11$~GHz) to the source of the transistor, effectively varying $V_\text{TG}$ at a fixed $V_\text{BG}$, as indicated along the set of lines in Fig.$\,$\ref{Figure2}(c). The MW field drives the system back and forth between the different charge states. For example, if the system begins in state (11), indicated by point A in Fig.$\,$\ref{Figure2}(c), and is then driven to lower gate voltage (point B), an electron exits the DQD: state (10). At even lower gate voltages (point C), the system traverses the (10)-(01) anticrossing, the system performs a LZSM transition and its wave function is therefore split into two components, acquiring different dynamical phases. Upon a sweep back to higher gate voltage, the system undergoes a second passage through the anticrossing, resulting in interference in the probabilities of the (10) and (01) states (point B again). Finally, the state (01) is projected by relaxation to the (11) state in point A and the cycle starts again. Since $\omega_\text{r} \ll \omega$, the resonator sees an average of the occupation probabilities of the DQD at each point in detuning. These changes in probabilities manifest as changes in the parametric capacitance of the DQD, which we detect via changes in the phase response.

In Fig.$\,$\ref{Figure2}(d), we plot the results of the drive sequence in Fig.$\,$\ref{Figure2}(c), where we show the normalized phase response, $\Delta\Phi/\Delta\Phi^0$, as a function of $V_\text{TG}$ for $V_\text{BG}=-4.3$~V. For $V_\text{TG}<0.4875~V$, where LZSM interference occurs, we observe the predicted oscillatory phase response. The oscillations decrease in amplitude when decreasing $V_\text{TG}$, as predicted by our model, see Fig..$\,$\ref{Figure1}(d). Finally, for $V_\text{TG}>0.4875~V$, the phase response decays rapidly as predicted by Eq.~(\ref{Cp_final}).


\section{Comparison between theory and experiment}

Comparing Fig.~\ref{Figure1}(d) and Fig.~\ref{Figure2}(d), we observe a good agreement between our theoretical prediction and the experiments. The calculations reproduce the experimentally observed sinusoidal dependence and amplitude attenuation of the capacitance, with good agreement in the voltage regions in which the LZSM experiments were performed. Now, we explore further the validity of our model by probing the system at different MW frequencies. In Fig.~\ref{Figure3}(a), we show the normalized resonator phase response as a function of $V_\text{TG}$ for six different frequencies, ranging from 4.72~GHz (black) to 21~GHz (yellow). Additionally, in Fig.~\ref{Figure3}(b), we show the normalized parametric capacitance obtained with Eq.~(\ref{Cp_final}) using the same frequencies as in the experiment, with $T_{1}=50$~ns, $T_2=35$~ps, and $T_\text{R}=30$~ps. We observe that our model reproduces well the experimental results. It captures the frequency and detuning dependence of the amplitude oscillations, as well as the change in oscillation lineshape at the highest MW frequencies (see yellow trace). Changing the rate at which the system is driven enables testing both the sinusoidal dependence and the amplitude of the signal predicted by Eq.~(\ref{Cp_final}).

\begin{figure}[htb]
\includegraphics{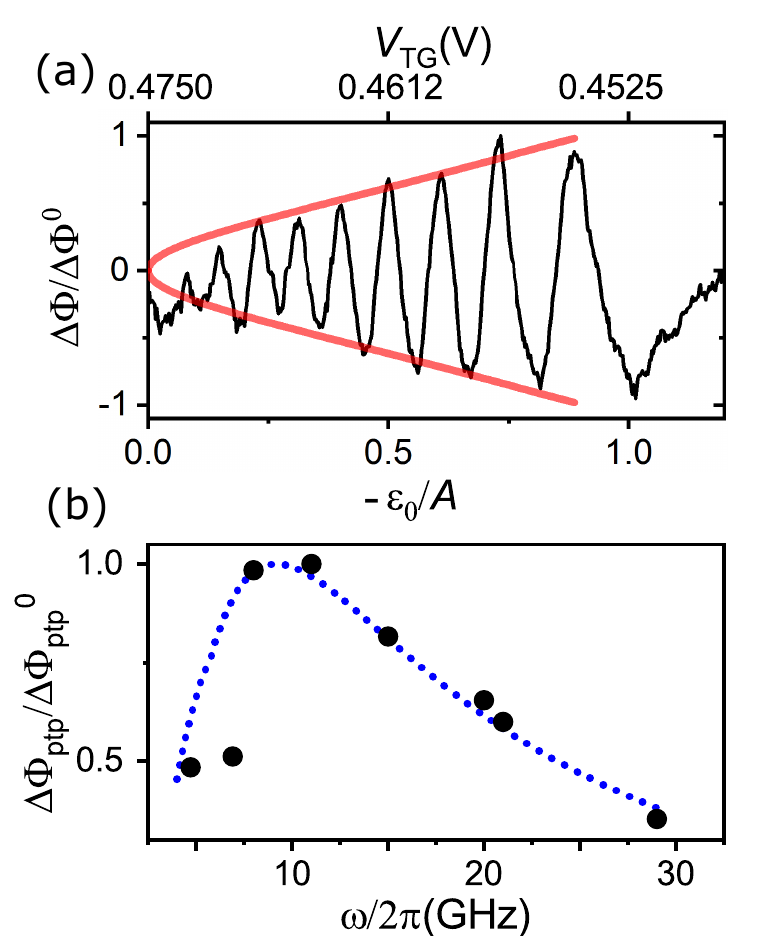}
\caption{\label{Figure4} Amplitude analysis. (a) Experimental normalized phase response as a function of reduced detuning for $\omega$/2$\pi=11$~GHz and the envelope of the oscillations obtained using the envelope of Eq.~($\ref{Cp_final}$) and $T_2=35$~ps and $T_\text{R}=30$~ps (red). (b) Experimental normalized peak-to-peak amplitude of phase response (black dots). Calculated normalized peak-to-peak parametric capacitance as a function of the MW frequency for an intrinsic relaxation time $T_{1}=50$~ns and $T_2=35$~ps and $T_\text{R}=30$~ps (blue dotted line).}
\end{figure}


First, we explore the sinusoidal shape of the signal. In Fig.~\ref{Figure3}(c), we plot the top-gate voltage at which the maximum in the Fourier transform of the data in Fig.~\ref{Figure3}(a) occurs for different MW frequencies. We observe a linear relation between both magnitudes. The results confirm the functional dependence between the parametric capacitance and $V_{\text{TG}}$ proposed in Eq.$\,$(\ref{MasterEq.Final_2}). From the fit we extract a QD gate coupling difference, $\alpha_-=0.06\pm 0.004$.

Next, in Fig.~\ref{Figure4}, we explore the dependence of the amplitude of the capacitance oscillations with $V_\text{TG}$ (or equivalently $\varepsilon_0$) and $\omega$. In Fig.~\ref{Figure4}(a), we show the data for the normalized phase response as a function of reduced detuning, where we have used the relation, 

\begin{equation}
\varepsilon=-2e\alpha_{-}\left(V_{\text{TG}}-V_{\text{TG}}^{0}\right),
\end{equation}

 with $V_{\text{TG}}^{0}=0.475$~V and $A=1.35$~meV. In this case, we show the data for the cycle involving (00)-(10) particle exchange with the reservoir to show the symmetry of the signal with respect to $\varepsilon=0$. Here, we observe the amplitude of the oscillations decaying with increasing $V_\text{TG}$. Looking at Eq.~($\ref{Cp_final}$), we see that the envelope of the oscillations is determined by $T_2$ and $T_\text{R}$. Intuitively, for values $\varepsilon_0\approx A$, the system spends less time after the first passage and hence the effect of decoherence in the amplitude of the signal is reduced. Additionally, at this detuning setting, the system has more time to tunnel to the reservoir increasing the overall amplitude of the signal. On the contrary, for values $\varepsilon_0\approx 0$, the system has more time to decohere and less to tunnel to the reservoir, leading to a reduced phase amplitude. The shape of the envelope allows determining $T_2=35$~ps and $T_\text{R}=30$~ps, extracted from the fit (red lines in Fig.~\ref{Figure4}(a)) .


Finally, in Fig.~\ref{Figure4}(b), we explore the peak-to-peak amplitude of the capacitance oscillations as a function of $\omega$. As we increase the frequency, we observe an increase in the peak-to-peak amplitude until $\omega/2\pi\approx 10$~GHz, where it starts to decay. These results can be explained as a competition between the different timescales of the system, $T_1$, $T_2$, and $T_\text{R}$, as can be seen in Eq.~(\ref{Cp_final}). In the following, we explain this competition qualitatively. Starting at low $\omega$, where the frequency is still comparable to the decoherence rate, the system cannot always complete the LZSM interference cycle leading to a lower capacitance signal. As we increase the frequency, the signal increases because, on average, more LZSM cycles are completed. However, as we continue increasing $\omega$, the $T_\text{R}$ processes start to matter since the system may not have sufficient time to relax to the reservoir. The position of the maximum in this experiment is determined primarily by the competition of these two processes. However, in general the ratio between the LZSM transition rate and the DQD relaxation rate influences the position of the maximum. In our case, we observe a dependence of the maximum with $T_1$ that enables estimating this parameter. In Fig.~\ref{Figure4}(b), we plot the best fit, using the already extracted values of $T_2=35$~ps and $T_\text{R}=30$~ps, and find $T_1=50$~ns. Both charge relaxation and coherence times are compatible with other measurements in silicon qubits~\cite{Dupont-Ferrier2013}. Overall, the good agreement between the theoretical model and the experiment indicates a viable scheme for the quantum interference capacitor and enables understanding the different timescales of the system from the shape of the capacitance curves.

\section{Conclusions}

In this article, we have introduced the idea of a capacitor that obtains its functionality from quantum interference in a system with discrete charge states. We have demonstrated a particular implementation using a single-electron double quantum dot coupled to an electron reservoir under the effect of a strong MW driving field. The system shows an oscillatory behaviour of the capacitance as a function of the QD energy level detuning, whose amplitude is determined by the charge relaxation time $T_1$, coherence time $T_2$, and tunneling time to the reservoir $T_\text{R}$. The voltage period of the capacitance oscillations is directly proportional to the frequency of the MW excitation.  Our model, based on a semi-classical master-equation formalism, captures the dynamics of the system and enables predicting the capacitive response of a DQD in the double-passage LZSM regime. Our work opens up an opportunity to use the non-linear capacitance of double quantum dots to design devices with enhanced functionality.

\section{Acknowledgments}

We thank K.~Ono for useful discussions. This research has received funding from the European Union's Horizon 2020 Research and Innovation Programme under grant agreement No.~688539 (http://mos-quito.eu). MFGZ acknowledges support from the Royal Society and the Winton Programme of the Physics of Sustainability. F.N. is supported in part by the MURI Center for Dynamic Magneto-Optics via the Air Force Office of Scientific Research (AFOSR) (FA9550-14-1-0040), Army Research Office (ARO) (Grant No.~W911NF-18-1-0358), Asian Office of Aerospace Research and Development (AOARD) (Grant No.~FA2386-18-1-4045), Japan Science and Technology Agency (JST) (Q-LEAP program and CREST Grant No.~JPMJCR1676), Japan Society for the Promotion of Science (JSPS) (JSPS-RFBR Grant No.~17-52-50023 and JSPS-FWO Grant No.~VS.059.18N), RIKEN-AIST Challenge Research Fund, and the John Templeton Foundation. AC acknowledges support from the EPSRC Doctoral Prize Fellowship. 

\nocite{apsrev41Control}

\bibliographystyle{apsrev4-1}
\bibliography{LZS_enhancement}
\end{document}